\documentstyle[amssymb,twocolumn,11pt]{article}

\begin{document}

\title{Beyond Unruh Effect: Nonequilibrium Quantum Dynamics of Moving Charges}
\author{B. L. Hu\thanks{%
Electronic address: {\tt hub@physics.umd.edu}} and Philip R. Johnson\thanks{%
Electronic address: {\tt pj19@umail.umd.edu}} \\
%EndAName
Department of Physics, University of Maryland\\
College Park,Maryland 20742-4111}
\date{December 1, 2000}

\maketitle

\begin{abstract}
We discuss some common misconceptions in Unruh effect \cite{Unr76} and Unruh
radiation for the cases of linear and circular uniform acceleration of a
charged particle or detector moving in a quantum field. We point to the need
to go beyond Unruh effect and develop a new theoretical framework for
treating the stochastic dynamics of particles interacting with quantum
fields under more general nonequilibrium conditions. This framework has been
established in recent years using the influence functional formalism \cite
{RHA,RHK,RavalPhD} and applied to relativistically moving charged particles 
\cite{JohnsonPhD,CapJH,JH}. Only with nonequilibrium concepts and
methodology applied to particle-field interaction can one grasp the full
complexity of the problems of beam physics under more realistic conditions,
from electrons and heavy ions to coherent atoms.
\end{abstract}

\noindent {\small {\it Invited talk given by BLH at the Capri Workshop on
Quantum Aspects of Beam Physics, Oct. 2000. To appear in the Proceedings
edited by Pisin Chen (World Scientific, Singapore, 2001).}}

\section{Introduction and Summary}

In this talk we would like to address two sets of issues, one related to
Unruh effect, the other related to moving charges in a quantum field, with
the hope of clarifying some misconceptions related to these problems. Unruh
effect attests that a detector (made of an oscillator, atom, electron, or
particle states of a quantum field) moving with a uniform proper
acceleration of magnitude $a$ sees the vacuum state of a quantum field as a
thermal bath with temperature $T_{U}=\hbar a/(2\pi ck_{B})$. This effect may
be understood purely as a kinematic aspect of ordinary quantum field theory
and does not require the notion of horizon, despite the connection with the
black hole Hawking effect \cite{Dalian}. It is important to recognize that
the Unruh effect is a manifestation of thermal noise in the detector, not
radiation from the detector. We explain this point below. The first set of
issues of interest are:

1) {\it Is there radiation emitted from a uniformly accelerated detector 
\cite{Chen}?} This is the title of the other talk by BLH, contained in a
summary paper by Hu and Raval in this volume \cite{CapHR}. The simple answer
is NO, when the detector has reached a steady state. There is emitted
radiation in nonequilibrium conditions associated with transients or
nonuniform accelerated motion (though the time for a uniformly accelerating
charge to equilibrate may be quite long). One example of nonequilibrium
conditions is finite time acceleration. This problem was treated with the
influence functional method by Raval, Hu and Koks \cite{RHK}. The other
example of nonequilibrium (though stationary) condition is the case of
circular motion, to which one can ask the question:

2) {\it Is there a circular Unruh effect \cite{BelLen}?} The strict answer
is NO, in the sense that the detector undergoing circular motion will NOT
detect a thermal bath, and hence there is strictly speaking no associated
Unruh temperature. Laboratory (e.g., storage ring) conditions may allow a
range of parameters (radius versus angular acceleration) such that a
near-equilibrium condition exists, in which case and only in such cases can
one use the concept of effective temperature, such as was proposed by Unruh 
\cite{Unr98}. Under general conditions, the moving particle/detector will
register a colored noise, (which turns white in linear uniform
acceleration), and acquire a stochastic component in its trajectory and
other degrees of freedom.

For treating these general cases, one needs to invoke statistical field
theory applied to the nonequilibrium dynamics of moving charges or detectors
in a quantum field. This is the subject matter of the Ph.D. theses of Alpan
Raval and Philip Johnson. A partial summary of the latter work, specifically
on the derivation of the Abraham-Lorentz-Dirac (ALD) equation \cite{ALD} and
its stochastic counterpart, the ALD-Langevin equation, is contained in our
other paper in this volume. To facilitate our discussion of this class of
problems, including the ``circular Unruh effect'', we need to develop some
basic concepts such as backreaction, fluctuations, dissipation and
decoherence, and understand the demarcation of quantum, stochastic and
semiclassical regimes. For this we bring in the second set of issues:

3) {\it Are radiation reaction (RR) and vacuum fluctuations (VF) related by
a fluctuation-dissipation relation (FDR)?} The answer is NO, not directly. 
{\it Is there a FDR at work?} YES. But it relates vacuum fluctuations to
quantum dissipation distinguished as the quantum backreaction which is over
and above the classical radiation reaction. It balances the stochastic
component in the particle trajectory so that the noise-averaged mean
trajectory follows a semi-classical equation of motion.

4) {\it Are runaway solutions and preacceleration necessary evils of ALD
equation?} NO, if one adopts the correct conceptual framework and
methodology. Key to the resolution of these puzzles is the concept of
decoherent history and emergent classical behavior from quantum systems.
Vacuum fluctuations not only bring about quantum dissipation, it is also a
source for decoherence in the quantum system. Decoherence legitimatizes a
classical description such as particle trajectories. We will discuss the
gist of these issues in the following sections. Full details can be found in
the original papers.

\section{Quantum, Stochastic, Semiclassical and Classical}

\subsection{Quantum Open System}

A closed quantum system can be partitioned into several subsystems according
to the relevant physical scales. If one is interested in the details of one
such subsystem, call it the distinguished {\it system}, and decides to
ignore certain details of the other subsystems, comprising the {\it %
environment}, the distinguished {\it system} is thereby rendered an
open-system. The overall effect of the coarse-grained environment on the
open-system can be captured by the influence functional technique of Feynman
and Vernon, or the closely related closed-time-path effective action method
of Schwinger and Keldysh \cite{ifctp}. These are initial value formulations.
For the model of particle-field interactions under study, this approach
yields an exact, nonlocal, coarse-grained effective action (CGEA) for the
particle motion \cite{cgea}. The CGEA may be used to treat the
nonequilibrium quantum dynamics of interacting particles. However, only when
the particle trajectories become largely well-defined (with some degree of
stochasticity caused by noise) as a result of decoherence due to
interactions with the field can the CGEA be meaningfully transcribed into a
stochastic effective action, describing stochastic particle motion. In this
program of investigation we take a microscopic view, using quantum field
theory as the tool to give a first-principles derivation of moving particle
interacting with a quantum field from an open-systems perspective.

\subsection{Fluctuation-Dissipation Relations}

A consequence of coarse-graining the (quantum field) environment is the
appearance of noise which is instrumental to the decoherence of the system
and the emergence of a classical particle picture. At the {\bf semiclassical}
level, where a classical particle is treated self-consistently with
backreaction from the quantum field, an equation of motion for the {\it mean}
coordinates of the particle trajectory is obtained. This is identical in
form to the classical equation in the case of linearly coupled theories.
Backreaction of radiation emitted by the particle on the particle itself is
called {\it radiation reaction}. (For the special case of uniform
acceleration it is equal to zero, due to a balance between the acceleration
field and the radiation field \cite{Jackson}.) Radiation reaction (RR) is
often regarded as balanced by {\it vacuum fluctuations} (VF) via a
fluctuation dissipation relation (FDR). This is a misconception: RR exists
already at the classical level, whereas VF is of quantum nature. There is
nonetheless a FDR at work balancing quantum dissipation (the part which is
over and above the classical radiation reaction) and vacuum fluctuations.
But it first appears only at the {\bf stochastic} level, when self
consistent backreaction of the {\it fluctuations} in the quantum field is
included in our consideration. Fluctuations in the quantum field is also
responsible for a stochastic component in the particle trajectory (beyond
the mean). Their balance is embodied in a set of generalized
fluctuation-dissipation relations.

\subsection{Decoherent Histories, Preacceleration and Runaway Solutions}

Not only can coarse-graining of the environment lead to dissipation in the
system dynamics, it is also responsible for the decoherence and emergence of
classicality in the system, such as the appearance of a classical
trajectory. When the environment is a quantum field and the system
decoheres, then quantum fluctuations can act effectively as a classical
stochastic noise \cite{HM2,CH94}.

The view that semiclassical solutions arise as decoherent histories \cite
{GelHar2} also suggests a new way to look at the radiation-reaction problem
for charged particles. The classical equations of motion with backreaction
are the Abraham-Lorentz-Dirac (ALD) equations. The solutions to the ALD
equations have prompted a long history of controversy due to such puzzling
features as pre-accelerations, runaways, and the need for higher-derivative
initial data \cite{Plass}. It has long been felt that the resolution of
these problems must lie in the progenitory quantum theory. But this still
leaves open the question of when, if ever, the ALD equation appropriately
characterizes the classical limit of particle backreaction; how the
classical limit emerges; and what imprints the correlations of the quantum
field environment leave. Further questions pertinent to the classical
behavior arising from the quantum realm, in the context of a moving charge
in a quantum field, include whether the decoherent histories are 1)
solutions to the ALD equation, 2) unique and runaway free, and 3) causal (no
pre-acceleration). In \cite{JH} we show how these puzzles and pathologies,
both technical and conceptual, are resolved in the context of the initial
value quantum open system approach, and that quantum corrected ALD equations
satisfying these criteria describe the semiclassical limit.

\section{ Radiation Reaction and Vacuum Fluctuations}

\subsection{Classical Radiation and Radiation Reaction}

Uniformly accelerated charges classically radiate according to the Larmor
formula, but experience vanishing RR \cite{Jackson}. There is an existing
belief that the extra work done on the charge against RR must be the direct
source of radiant energy, but this static viewpoint is inappropriate. Fields
are dynamical objects and have complex interactions with particles. For
example, the acceleration field has been shown to do work on charges (and
visa versa) and therefore one can not expect a detailed balance between
particle and radiation energy alone since that would require a ``freezing''
out of the near and intermediate field degrees of freedom in a way
incompatible with locality and causality.

\subsection{Quantum Radiation and Vacuum Fluctuations}

Let us now examine the quantum properties of this system. Our result based
on self-consistent backreaction says that the stochastic equations when
averaged over the noise distribution (noise-average) gives the (mean-field)
semiclassical form. In the uniform acceleration case with linear coupling
the expectation value of the field (quantum mean) is exactly the same as the
classical value where the particle/detector is treated as a ``classical''
source, though the mean particle trajectory must be self-consistently
determined as we have emphasized. At the stochastic level, the particle
detector does fluctuate in its worldline, and other degrees of freedom. How
does this stochastic component affect the field? As shown by Ravel, Hu and
Anglin \cite{RHA} (for an alternative derivation, see \cite{CapHR}),
fluctuations in a detector modify the near field correlations-- a
polarization cloud is found around the detector trajectory. The same is true
for stochastic particle motion in the linearized regime. This quantum effect
of modified field correlations adds on to the average classical field value
(the two-point function is different from the free field value). By
extrapolating the RHA results to 3+1 dimensions, one may see that these
altered field correlations showing up as vacuum polarization drop off faster
than $1/r^{2}$ and hence are not seen by observers at infinity \cite{TRH}.
Since the equivalence of a quantum mean to the classical value holds only
under the one-loop, Gaussian approximations, when these conditions are
lifted, there may be new effects as yet undiscovered.

Whether there is quantum-corrected radiation from a nonuniformly accelerated
charge or detector is therefore what one should focus on here when one asks
a question like ``Is there emitted radiation in Unruh effect?'' Our result
obtained with self-consistent backreaction of quantum fluctuations shows
that the (noise-averaged) of a decohered particle trajectory obeys the ALD
equation, which is known to be consistent with the classical Larmor formula
(if one include the nonlocal acceleration field effects, as one must). This
applies to any accelerated trajectory, uniform or nonuniform, which implies
that there is no additional ``extra'' average radiation in the
semiclassical/stochastic regime beyond the usual classical quantity, even
though there are fluctuations (noise) induced in the particle (the Unruh
effect in the uniform acceleration case). It has been verified that the
presence of detector fluctuations is not inconsistent with the absence of
additional radiation.

When quantum decoherence is incomplete, the mean-field equations of motion
for both radiation and particle have quantum corrections (an example of this
is Schwinger's synchrotron radiation calculation \cite{Schwinger}) which
must be included to answer questions beyond the semiclassical or stochastic
domain.

\subsection{Nonequilibrium quantum dynamics of charges}

One major improvement of our approach to the problem of moving charges in a
quantum field is the consideration of full backreaction of the quantum field
on the particle in the determination of its trajectory. Dynamical
backreaction ensures self-consistency between the particle/detector and the
quantum field. The lack thereof is where many of the problems and paradoxes
arise. We also find that conceptual issues are easier to consider if we deal
with such problems at four distinct levels: quantum, stochastic,
semiclassical and classical, as explained earlier. Confusion will arise when
one mixes physical processes of one level with another without knowing their
interconnections, such as drawing the equivalence between radiation reaction
with vacuum fluctuations. Before summarizing our thoughts for processes
under nonequilibrium conditions, which cover most cases save a few special
yet important ones, such as uniform acceleration, let us remark that these
well-known cases are what we would call `test field' or prescribed
(trajectory) cases and not self-consistent or backreaction-sensitive. These
cases are easier to study because they possess some special symmetry, such
as is present for the uniform acceleration case (Rindler spacetime),
inertial case (Minkowski), or the eternal black hole case (Killing tensor).
They are legitimate only if the backreaction of the field on the particle
permits such solutions. Under these special conditions, a detector feels a
thermal bath (in the inertial case it is the zero-temperature vacuum).

Let us analyze the physics of nonequilibrium processes at separate levels:

{\bf Classical} level- the decohered self-consistent (mean) solutions for
particle and field. If the system is sufficiently coarse-grained and
decohered, the particle obeys classical equations of motion, such as the ALD
equation from QED \cite{CapJH}. There is no Unruh effect because it is
quantum in nature (at the classical level the effect of quantum fluctuations
are averaged out).

{\bf Semiclassical} level -- defined as a classical system (particles or
detectors) interacting with a quantum field. Coarse-graining over quantum
field for reduced particle dynamics at one-loop gives back the classical
equations of motion for the mean trajectory of the particle. Higher-order
quantum corrections arising from nonlinearities modify the mean of the
quantum equations of motion for the particle. Quantum corrections may not
however show up significantly at the low energy macroscopic description
because decoherence tends to suppress these higher-order (e.g., higher-loop)
nonlinear quantum effects.

{\bf Stochastic} level - where fluctuations of the quantum field manifest as
stochastic noise in the system dynamics. Coarse-graining the field (to some
but not the fullest --classical --extent), one obtains a classical
stochastic equation for the system (such as the Einstein Langevin equation
for semiclassical stochastic gravity \cite{stogra,MarVer} or the
ALD-Langevin equation for QED \cite{CapJH,JH}). It is possible to encode
much of the quantum statistical information of the field and the state of
motion of the system in the noise correlator and the two point function of
the particle. Thus effects of both quantum (field environment) and kinematic
(particle system) nature show up as a stochastic component in the particle
trajectory which is self-consistently determined. The stochastic equations
of motion have a quantum dissipation term (not classical radiation
reaction!) that balances the quantum fluctuations, and is governed by a FDR.
The latter is described by the noise kernel, which for general conditions is
nonlocal, entailing that the noise in the detector is colored and
temperature is no longer a viable concept.

\section{`Circular Unruh Effect' -- Misconceptions}

We now apply these ideas to discuss radiation from a particle in circular
motion in a quantum field and in particular we address two common sets of
misconceptions related to it. (We only present the main points here, see 
\cite{HJT} for calculations and further discussions.) These misconceptions
arise from unclear distinction between a) linear uniform acceleration and
circular motion, b) thermal radiance felt by the detector/charge in uniform
acceleration (Unruh effect) versus emitted radiation (misconjured as Unruh
`radiation') sensed by probes afar, and c) emitted radiation of classical
and quantum origin.

It has been asserted that Unruh radiation is already observed in storage
rings \cite{BelLen}. This is the so-called circular Unruh effect. For this
discussion we assume that RF fields give the particle average circular
(steady state) motion by restoring the energy loss from synchrotron
radiation. Questions:

\subsection{Is there a circular Unruh effect?}

NO. In fact, the circular case displays nonequilibrium (albeit steady state)
quantum field statistics that are more general than the linear uniform
(thermal) Unruh case. There is a difference between linear acceleration and
angular acceleration. Just from dimensional grounds, there is only one
parameter in the linear case, the proper acceleration $a$, but two in the
circular case, the angular acceleration $\alpha $ and the radius of the
orbit $R$. In the linear case, as the velocity of the particle increases to
the speed of light, an event horizon forms. In the circular case, the
direction of velocity changes but its magnitude remains constant, there is
no event horizon. (Invoking Kerr metric to describe circular motion is
unnecessary and misleading, as the problem is basically about kinematics in
relativistic quantum field theory.)

\subsection{Is temperature a viable concept?}

NO. To the extent that the existence of an event horizon is the condition
for the appearance of an Unruh or Hawking temperature (this is the
traditional argument based on global geometry \cite{HarHaw}, the modern one
is via kinematic effect, which enables one to consider nonequilibrium
conditions \cite{Dalian}), one can already see that there is no well-defined
Unruh temperature in circular motion. For circular motion one needs to
incorporate the effect of a second physical scale other than acceleration
(e.g., the radius). If the system is in near-equilibrium conditions, one can
introduce an `effective (frequency dependent) temperature' \cite{Unr98}.

\subsection{Emittance and Vacuum fluctuations}

A related point is the emittance (spread) of particle beams, which is
commonly understood to result from quantum field-induced fluctuations. One
can treat beam emittance without invoking temperature or Unruh effect. For
general cases there is no need for temperature to play the intermediary
between quantum field and induced beam fluctuations (on this point we concur
with Jackson \cite{Jackson98}).

Beam emittance is indeed the working of kinematic effects (particle motion)
on vacuum fluctuations (quantum noise). (For viewing Hawking -Unruh effect
in this light see \cite{Dalian}). Beams in linear uniform acceleration are
expected to show thermal spread (neglecting possible sources of non-thermal
noise). Beams in circular motion do not come into thermal equilibrium,
though they may achieve a steady state balance between vacuum fluctuations
and quantum dissipation. Our prediction is that the detector (a particle
with internal degrees of freedom such as an electron with spin) will see
colored noise whose correlator is related to the nonthermal electron
populations in their two polarization states. This is more general than the
Unruh effect as it is under nonequilibrium conditions.

\subsection{Isn't synchrotron radiation Unruh radiation?}

No. Synchrotron radiation occurs for classical systems (where there is no $%
\hslash $); or arises in the semiclassical limit of quantum systems where
quantum noise has been averaged out. The Unruh {\it effect }is thermal
radiance in the system arising from quantum fluctuations; it is seen in the
stochastic and quantum limit. One argument views synchrotron radiation as
the scattering of virtual vacuum fluctuations into real photons by a moving
charge. But in the Unruh effect there is no radiation after the system has
equilibrated, yet there are thermal fluctuations in the particle. This
highlights the distinction between emitted radiation (synchrotron or Larmor)
and thermal radiance felt by the particle/detector (Unruh effect). There is
no direct link between the classical limit of radiation and the quantum
Unruh effect; but at the stochastic level a FDR\ relates quantum dissipation
and vacuum fluctuations \cite{JH}.

\subsection{Is there emitted quantum radiation from the charge?}

At the stochastic level there is nonequilibrium noise in the
particle/detector; these fluctuations alter field correlations around the
particle trajectory as a polarization effect \cite{RHA}. At the quantum
level one can use the open system approach but coarse-grain the particle,
and determine the quantum corrections to radiation. Take note that quantum
corrections modifying both the mean-field radiation and noise-average
trajectory must be found self-consistently. The result should be compared
with Schwinger's \cite{Schwinger} and/or the quasi-classical operator method
because discrepancies, if any, will be of considerable interest.

\section*{Acknowledgments}

We thank Pisin Chen for his invitation to this interesting workshop and
Stefania Petracca for her warm hospitality. This research is supported in
part by NSF grant PHY98-00967 and DOE grant DEFG0296ER40949.


\begin{thebibliography}{99}
\bibitem{Unr76}  W. G. Unruh, Phys. Rev. {\bf D14}, 3251 (1976).

\bibitem{RHA}  A. Raval, B. L. Hu and J. Anglin, Phys. Rev. {\bf D 53}, 7003
(1996).

\bibitem{RHK}  A. Raval, B. L. Hu and D. Koks, Phys. Rev. {\bf D 55}, 4795
(1997).

\bibitem{RavalPhD}  A. Raval, Ph. D. Thesis, University of Maryland, College
Park, 1996.

\bibitem{JohnsonPhD}  P.~R. Johnson, Ph.D. thesis, University of Maryland,
College Park, 1999.

\bibitem{CapJH}  P.~R. Johnson and B.~L. Hu, ''Worldline Influence
Functional Derivation of Lorentz- Dirac- Langevin Equation from QED ''. This
volume.

\bibitem{JH}  P.~R. Johnson and B.~L. Hu, {\it Stochastic Theory of
Relativistic Particles Moving in a Quantum Field: I, II, III} (Preprints
2000).

\bibitem{Dalian}  B. L. Hu, ``Hawking-Unruh Thermal Radiance as Relativistic
Exponential Scaling of Quantum Noise'', Invited talk at the Fourth
International Workshop on Thermal Field Theory and Applications, Dalian,
China, August, 1995. Proceedings edited by Y. X. Gui and K. Khanna (World
Scientific, Singapore, 1996) gr-qc/9606073.

\bibitem{Chen}  Pisin Chen, ``Event Horizon'' This volume.

\bibitem{CapHR}  B. L. Hu and A. Raval, ``Is there Radiation in Unruh
Effect?''. This Volume.

\bibitem{BelLen}  J. M. Lennias, "Unruh Effect in Storage Rings" This volume.

\bibitem{Unr98}  W. G. Unruh, in Monterey Workshop on Quantum Aspects of
Beam Physics, edited by Pisin Chen (World Scientific, Singapore, 1998).

\bibitem{ALD}  H.~A. Lorentz, {\it The Theory of Electrons }(Dover Books,
New York, 1952), pp.\ 49,253; P.~A.~M. Dirac, Proc. R. Soc. London {\bf A 167%
}, 148 (1938).

\bibitem{ifctp}  R. Feynman and F. Vernon, Ann. Phys. {\bf 24}, 118 (1963);
J. Schwinger, J. Math. Phys. {\bf 2} (1961) 407; L. V. Keldysh, Zh. Eksp.
Teor. Fiz. {\bf 47 }, 1515 (1964) [Engl. trans. Sov. Phys. JEPT {\bf 20},
1018 (1965)].

\bibitem{cgea}  B. L. Hu and Y. Zhang, ``Coarse-Graining, Scaling, and
Inflation" Univ. Maryland Preprint 90-186 (1990); B. L. Hu, in {\it %
Relativity and Gravitation: Classical and Quantum} Proc. SILARG VII,
Cocoyoc, Mexico 1990. eds. J. C. D' Olivo et al (World Scientific, Singapore
1991).

\bibitem{Jackson}  J. D. Jackson, {\it Classical Electrodynamics} (J. Wiley,
N. Y. 1983).

\bibitem{HM2}  B. L. Hu and A. Matacz, Phys. Rev. D49, 6612 (1994).

\bibitem{CH94}  E. Calzetta and B. L. Hu, Phys. Rev. {\bf D49}, 6636 (1994).

\bibitem{GelHar2}  M. Gell-Mann and J. B. Hartle, Phys. Rev. {\bf D47}, 3345
(1993).

\bibitem{Plass}  G. N. Plass, Rev. Mod. Phys. {\bf 33}, 37 (1961); F.
Rohrlich, Am. J. Phys. {\bf 28}, 639 (1969).

\bibitem{TRH}  M. L. Tseng, A. Raval and B. L. Hu, in preparation (2000).

\bibitem{Schwinger}  J. Schwinger, Proc. Nat. Acad. Sci., {\bf 40}, 132
(1954); V. M. Baier and V. M. Katkov, {\it Zh. Eksp. Teor. Fiz.,} {\bf 53},
1478 (1967). transl., Sov. Phys. JETP 26, 854 (1968).

\bibitem{stogra}  B. L. Hu, Int. J. Theor. Phys. 38, 2987 (1999)
gr-qc/9902064

\bibitem{MarVer}  R. Mart\'{i}n and E. Verdaguer, Int. J. Theor. Phys. 38,
3049 (1999) .gr-qc/9812063. Phys.~Rev.~{\bf D60}, 084008 (1999).
gr-qc/9904021.

\bibitem{HJT}  B. L. Hu, P. R. Johnson and M-L Tseng (in progress).

\bibitem{HarHaw}  J. B. Hartle and S. W. Hawking, Phys. Rev. D13, 2188
(1976).

\bibitem{Jackson98}  J. D. Jackson, in Monterey Workshop on Quantum Aspects
of Beam Physics, edited by Pisin Chen (World Scientific, Singapore, 1998).

\bibitem{DDM}  P. Davies, T. Drey and C. Manogue, Phys. Rev. D (1998).
\end{thebibliography}
\end{document}